Elasticity of randomly cross-linked networks in primitive chain network simulations

Yuichi Masubuchi*

Department of Materials Physics, Nagoya University, Japan

*To whom correspondence should be addressed

mas@mp.pse.nagoya-u.ac.jp

ABSTRACT

Primitive chain network simulations for randomly cross-linked slip-link networks were performed. For the percolated networks, the stress-strain relationship was compared to the theories by Ball et al. [Polymer, 22, 1010 (1981)] and Rubinstein and Panyukov [Macromolecules, 35, 6670 (2002)]. The simulation results were reasonably reproduced by both theories, given that the contributions from cross-links and slip-links were used as fitting parameters. However, these parameters were model dependent. Besides, the theories cannot describe the simulation results if the parameters were determined from the number of active links involved in the percolated networks. These results reveal that the fitting of experimental data to the theories does not provide a fraction of entanglements in the system unless the network only consists of Gaussian strands and it correctly reaches the state of free-energy minimum.

INTRODUCTION

Contributions of entanglements to the elasticity of network polymers is one of the fundamental problems in polymer science[1]. In the earlier theories, so-called affine network models[2,3] and phantom network models[4,5], entanglements among polymers were not considered. Ronca and Allegra[6] implemented the effect of entanglement via the reduced fluctuations around the network node. For this approach, the so-called constrained-junction model, there have been several attempts in parallel[7,8]. Edwards[9] considered that entanglement restricts fluctuations of the subchains between network nodes by the tube-shaped constraint. Rubinstein and Panyukov[10] considered the fluctuations both for the network nodes and strands to propose the nonaffine tube model. In the theories mentioned above, entanglement between polymers is not explicitly considered, but its effect is embedded into the reduced fluctuations in the network.

In a few theories, the contributions from entanglements and cross-links are separately considered. Ball et al.[11] proposed the slip-link model, where the entanglement between polymers is replaced by a slip-link. This slip-link bundles two chains, and it slides along the connected chains. Using the replica formalism, they derived the free energy under stretch with a parameter that describes the amount of chain sliding at the slip-link. Rubinstein and Panyukov[12] implemented the slip-link idea to their nonaffine tube model to propose the slip-tube model. They elaborated to obtain the magnitude of sliding as a function of stretch ratio. Both theories can reproduce experimental data quantitatively if the model parameters are chosen to fit the data.

Despite the success, evaluation of the theories mentioned above is still necessary. Apart from the finite chain extensibility parameter[13,14], the fundamental parameters are the elastic moduli attributable to the contributions from cross-links and slip-links, $G_c$ and $G_e$. The other important parameter for the theory by Ball et al.[11] is the slippage parameter $\eta$ to describe the amount of chain sliding at the slip-link. In contrast, Rubinstein and Panyukov[12], theoretically determined the amount of slippage, and there is no parameter for the slippage. Nevertheless, as demonstrated later, the values of $G_c$ and $G_e$ for the fitting depend on the theory. This uncertainty of $G_c$ and $G_e$ means that the numbers of cross-links and slip-links may not correspond to those considered in the theories.

However, the evaluation is not straightforward because the experimental determination of the amount of entanglement is challenging. The active network strands carrying the stress must be discriminated from the free chains and dangling segments. For such a purpose, Oberdisse et al.[15,16] have performed coarse-grained simulations. They constructed some networks consisting of cross-links and slip-links by the end-linking reaction of monodisperse linear polymers. In their system, the number of slip-links on the chain between two consecutive cross-links is uniform, as assumed in the theories. By introducing a biased-sampling technique, they realized Gaussian chain statistics for the chains trapped in the network before the stretch. They reported that the magnitude of chain sliding at the slip-links is smaller than that expected from the theory by Ball et al.[11] Although the reason for this discrepancy has not been specified, it might be due to their network structure.

In this study, multi-chain slip-link simulations were conducted for randomly cross-linked networks. The employed model is basically the same as that used by Oberdisee et al.[15,16], except the network connectivity. Namely, after sufficient equilibration of entangled polymer melts, a fraction of slip-links was randomly switched to cross-links. The partially cross-linked systems were uniaxially stretched, and the stress-strain relationship was obtained. The simulation results exhibited that the amount of chain slippage at the slip-link was consistent with the theory by Ball et al.[11] Meanwhile, the contribution from entanglements to the network elasticity cannot be correctly estimated via fitting of the theoretical predictions to experimental data. Details are shown below.

MODEL AND SIMULATIONS

The model and the simulation code used in this study are the same as those employed in the previous studies for polymer melts under elongation[17–20], except the extension to cross-linked networks. In the model[21–23], an entangled polymeric liquid is replaced by a network consisting of network nodes, strands, and dangling ends. Each polymer chain in the system corresponds to a path connecting dangling ends through the strands and nodes. At each node, a slip-link is located to bundle two polymer chains. The slip-link allows the sliding of polymer chains along its backbone. The dynamics of the system are described by the motion of network nodes, the chain sliding, and the creation and destruction of the nodes around the chain end. The Langevin-type equation of motion describes the motion of network nodes according to the force-balance among the strand tension, drag force, osmotic force, and thermal random force. The chain sliding is represented by the transport of Kuhn segments between consecutive strands along the polymer backbone. The transport equation takes account of the same force-balance with the node motion. As a result of the dynamics, some dangling ends slide off or protrude from the connected slip-link. In the case of sliding-off, the slip-link is removed. Vice versa, if the number of Kuhn segments on a dangling end exceeds a certain maximum, a new slip-link is created on the dangling segment to hook another segment from the surroundings randomly. The model has been extended to branch polymers with the implementation of the arm retraction and the branch point withdrawal[18,24–28]. The calculated polymer dynamics and the resultant rheology are consistent with experiments for various entangled polymers, as reported previously.

In this study, the model was applied to network polymers, in which fractions of slip-links were randomly switched to cross-links for equilibrated melts of linear polymers. At the cross-links, the chain sliding was disallowed, and hence, no constraint release occurs. This cross-linked system was uniaxially stretched with the Poisson's ratio of 0.5 instantaneously, and the stress relaxation was observed. For the percolated networks, the stress does not decay to zero and mitigates to a steady value. The stress-strain relation was obtained for this constant stress to be compared with the theoretical predictions.

The simulations were conducted for prepolymers with the molecular weight of $Z=10$, 20, and 40, where $Z$ is the average number of network strands per chain. After equilibration of the prepolymer melts, a fraction $\varphi_c$ of slip-links changed into cross-links. Unit of length, energy, and time was chosen as the average strand length under equilibrium $a$, thermal energy $kT$, and the diffusion time of single node $\zeta a^2/kT$. Periodic boundary conditions were used with the simulation box size of $20^3$. The strand density was fixed at 10. The finite chain extensibility was not considered for simplicity. For each condition, four independent simulation runs starting from different initial configurations were performed for statistics.

RESULTS AND DISCUSSION

Figure 1 shows the fractions of active cross-links $\varphi_{ac}$ and trapped slip-links $\varphi_{as}$ in the system as functions of the nominal cross-linked fraction $\varphi_c$ for various prepolymer molecular weights. Here, $\varphi_{ac}$ and $\varphi_{ac}$ are the fraction of links involved in the percolated network, except the links on the dangling chains. Note that for small $\varphi_c$ because the system did not percolate $\varphi_{ac} = \varphi_{as} = 0$. Note also that for $\varphi_c = 1$, $\varphi_{as} = 0$ by definition. In comparison to the ideal behavior, where $\varphi_c = \varphi_{ac}$ shown by the broken line, $\varphi_{ac}$ is smaller than $\varphi_c$, reflecting the number of cross-links that are used for connecting isolated chains. The number of such inactive cross-links decreases with increasing $Z$ being consistent with the standard gelation theory. Meanwhile, $\varphi_{as}$ shows a maximum around $\varphi_c = 0.2$. This non-monotonic trend is related to the simulation setup as explained as follows. For small $\varphi_c$ cases, the trapped entanglement increases with increasing $\varphi_c$, reflecting an increase in the number of segments located between active cross-links. In contrast, for large $\varphi_c$, the trapped entanglements are switched to cross-links in the simulation setup. In other words, the maximum for $\varphi_{as}$ is partly due to the constraint imposed on the present model, in which the sum of cross-links and slip-links was unchanged. Nevertheless, for evaluation of the theoretical models, the obtained values of $\varphi_{ac}$ and $\varphi_{as}$ shall be employed later.

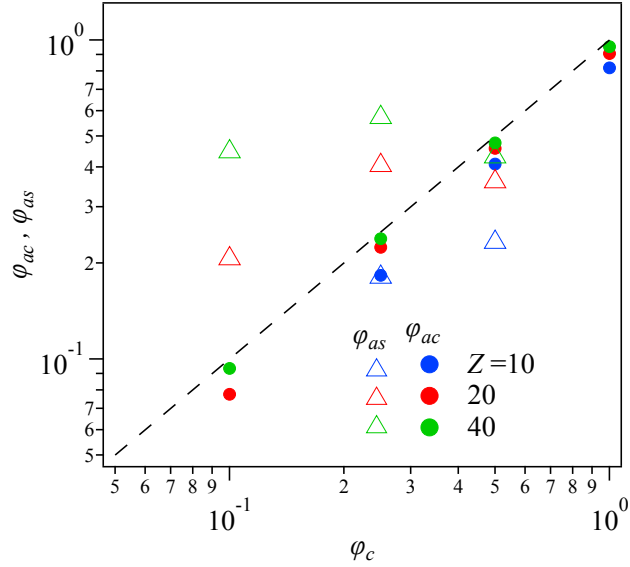

**Figure 1** Fraction of active cross-links $\varphi_{ac}$ (circles) and active trapped slip-links $\varphi_{as}$ (triangles) plotted against nominal cross-linking fraction $\varphi_c$ for various prepolymer molecular weights. The broken line shows the ideal behavior of $\varphi_{ac} = \varphi_c$.

Figure 2 shows the stress relaxation after an instantaneous uniaxial deformation with the Hencky strain of 3 (that corresponds to the stretch ratio of 9.9). For the cases with $\varphi_c = 0$, the stress decays to zero showing a two-step relaxation, reflecting the chain contraction followed by the orientational relaxation[29]. The relaxation time significantly increases with increasing $Z$, as established for entangled polymers. As $\varphi_c$ increases, the relaxation is slowed down due to the cross-linked chains. When $\varphi_c$ is small, the network is not percolated, and the stress goes to zero. Beyond a certain cross-linking fraction, the system became a solid, in which paths of infinite length made of the cross-linked chains exist. The stress does not decay to zero for the percolated networks, and it mitigates to a steady value.

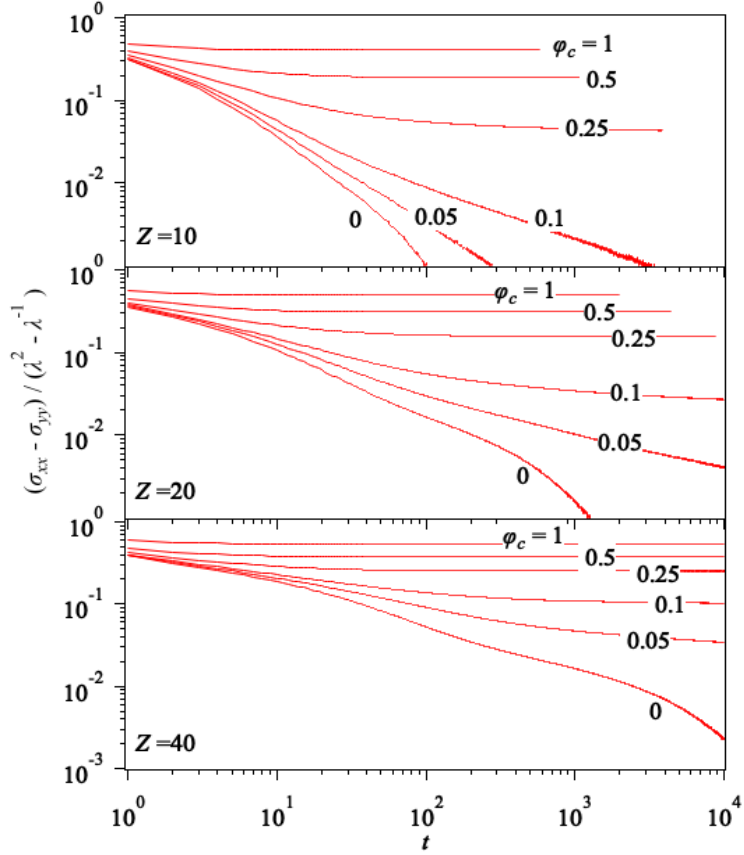

**Figure 2** Stress relaxation after an instantaneous uniaxial stretch with the Hencky strain of 3 for various nominal cross-linking fraction $\varphi_c$. The molecular weight of the prepolymer is 10, 20, and 40 from top to bottom.

Figures 3 and 4 shows the stress-strain relation in the Mooney-Rivlin plot[30,31] for the percolated networks in comparison to the theories by Ball et al. [11] (Fig. 3) and by Rubinstein and Panyukov[12] (Fig. 4). For the case with $\varphi_c = 1$ shown by black squares, in this plot, the stress exhibits no strain dependence, being consistent with the theories. Ideally, the value of Mooney stress for this case is $\varphi_{ac}/2$, according to the phantom network theory. However, the simulation results are higher than this theoretical value when $Z$ is large.

This discrepancy is due to the construction of the examined networks, which were produced from the snapshot of entangled polymers. As earlier reported, the number distribution of Kuhn segments on each entanglement strand is the exponential decay function, which has a large number of strands with zero Kuhn segment. Because the strand tension is given by the strand vector divided by the number of Kuhn segment, the tension generated by such strands diverges to infinity. Although a numerical cut-off was introduced for the Kuhn segment number on each strand, the strands that contain the minimum number of Kuhn segments emanate higher tension than the Gaussian spring. Such non-Gaussian strands induce the discrepancy in the macroscopic stress. Meanwhile, for small $Z$ cases, some dangling segments cause softening via additional fluctuations to the diverging cross-links. As a result, the simulation result obtained for $Z = 5$ is consistent with the theory.

For the network consisting of both cross-links and slip-links (i.e., $0 < \varphi_c < 1$), the modulus decreases with increasing the stretch due to the chain sliding. Both theories can describe this stretch-softening if the model parameters $G_c$ and $G_e$ are optimized irrespective of $\varphi_{ac}$ and $\varphi_{as}$ values, as shown by broken curves. However, if $G_c$ and $G_e$ are chosen to be consistent with $\varphi_{ac}$ and $\varphi_{as}$, the theories do not correctly reproduce the simulation results as demonstrated by solid curves. Note that the conversion of $\varphi_{ac}$ and $\varphi_{as}$ to $G_c$ and $G_e$ was according to the earlier study[12] as $G_c = \varphi_{ac}/2$ and $G_e = 4\varphi_{as}/7$ in the dimensionless form.

The comparison to the theory by Ball et al. [11] is shown in Fig 3. The solid curves underestimate the stress for large $Z$ values shown in the top and mid panels. However, the magnitude of stretch-softening is reasonably reproduced. The discrepancy from the simulation data can be mostly compensated if the value of $G_c$ is modified according to the case with $\varphi_c = 1$. This result demonstrates that the theory reasonably describes the simulated amount of slippage. As mentioned above, in the earlier simulation by Oberdisee et al. [15,16] for the end-linked networks, the amount of slippage was smaller. Their assumption for the uniform number of slip-links between consecutive cross-links may suppress the sliding. Nevertheless, the exact reason is unknown.

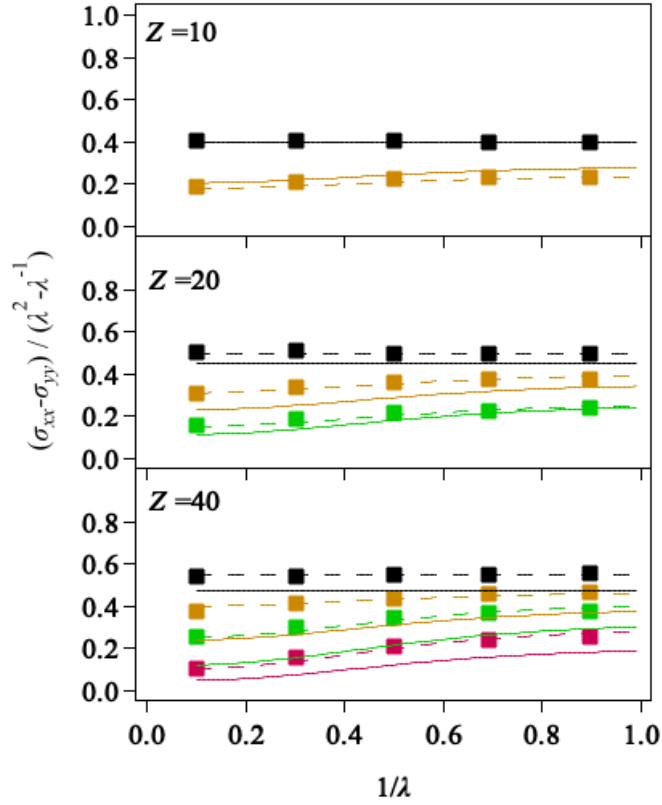

**Figure 3** Mooney-Rivlin plot of elongational stress for $Z$=40, 20 and 10, from top to bottom at $\varphi_c$=1 (black), 0.5 (orange), 0.25 (green), and 0.1 (magenta) in comparison to the theory by Ball et al. [11] Solid and broken curves are the theoretical predictions with the parameters determined from $\varphi_{ac}$ and $\varphi_{as}$, and optimized for the best fit, respectively. The sliding parameter is fixed at $\eta = 0.2$.

Figure 4 shows the comparison to the theory by Rubinstein and Panyukov. [12] The solid curves were drawn according to $\varphi_{ac}$ and $\varphi_{as}$. The results demonstrate that, for most cases, the predicted magnitude of stretch-softening is larger than the simulation. The contribution of the slip-springs under small stretch is overestimated. These results exhibit that the amount of chain sliding is smaller in the simulations, and the fluctuations around the slip-link are more extensive than those assumed theoretically. Because the theoretical slippage was derived from the free-energy minimization, the state of the stretched network does not correspond to the free-energy minimum. This idea can be visualized by the polymer chain, as shown in Fig 5. Here, a typical chain in the network is shown for $Z=40$ and $\varphi_c=0.5$. This chain is involved in a clustering structure of slip-links and cross-links seen on the right-hand side. The slip-links cannot slip freely in the cluster because the other links, including cross-links, restrict their motion. Meanwhile, for the case with $Z=40$ and $\varphi_c = 0.1$, the theory well-describes the simulation result. See magenta square and solid curve. In this case, the sliding sufficiently occurs, and the system can be close to the state with the minimum free-energy.

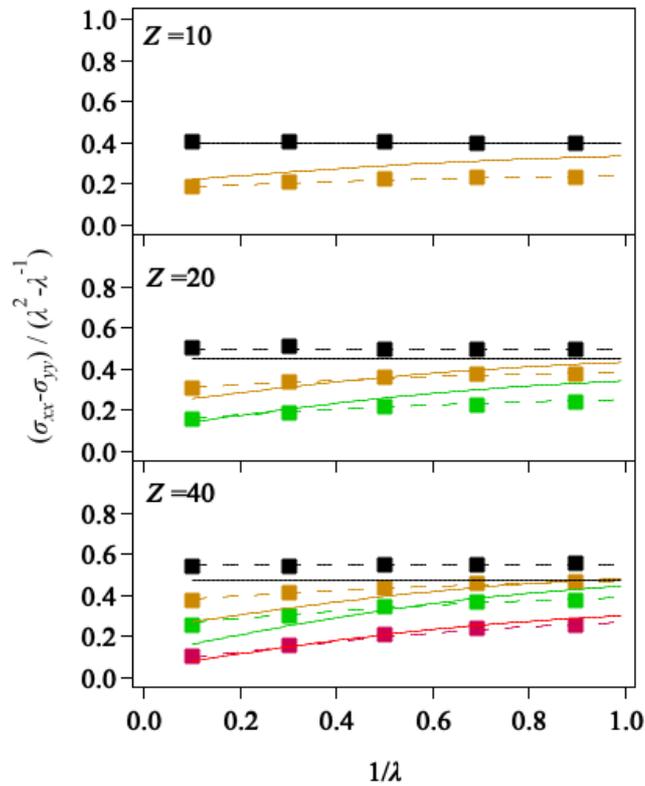

**Figure 4** Mooney-Rivlin plot of elongational stress for $Z=40$, 20 and 10, from top to bottom at $\varphi_c=1$ (black), 0.5 (orange), 0.25 (green) and 0.1 (magenta) in comparison to the theory by Rubinstein and Panyukov[12]. Solid and broken curves are the theoretical predictions with the parameters determined from $\varphi_{ac}$ and $\varphi_{as}$, and optimized for the best fit, respectively.

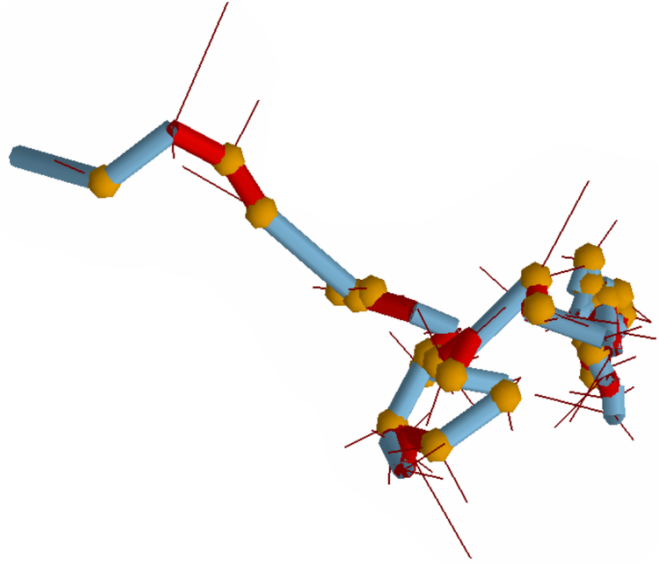

**Figure 5** Typical snapshot of polymer chain embedded in the percolated network for $Z=40$, $\varphi_c=0.5$, and $\lambda = 3.3$. Thin lines are the partner segments connected at cross-links and slip-links. The yellow ball shows the cross-links. The segments in blue and red carry the tension below and above a specific critical value.

Figure 6 shows the values of $G_c$ and $G_e$ that attain the best fit to the simulation data (see broken curves in Figs 3 and 4) as functions of $\varphi_{ac}$ and $\varphi_{as}$. Dotted lines show the relations written as $G_c = \varphi_{ac}/2$ and $G_e = 4\varphi_{as}/7$. For $G_c$ shown in the top panel, as mentioned for Figs 3 and 4, due to the non-Gaussian tension emanated by the network strands carrying a minimum number of Kuhn segments, the reasonable fitting was attained with the parameter values higher than the theoretical ones from the phantom network theory. Nevertheless, with a convex trend, the value of $G_c$ has a reasonable correlation with $\varphi_{ac}$ irrespective of the employed theory. In contrast, for $G_e$ shown in the bottom panel, poor correlation is observed with $\varphi_{as}$. This result implies that, at least for the simulated networks, the contribution of slip-links cannot be correctly evaluated by the examined theories.

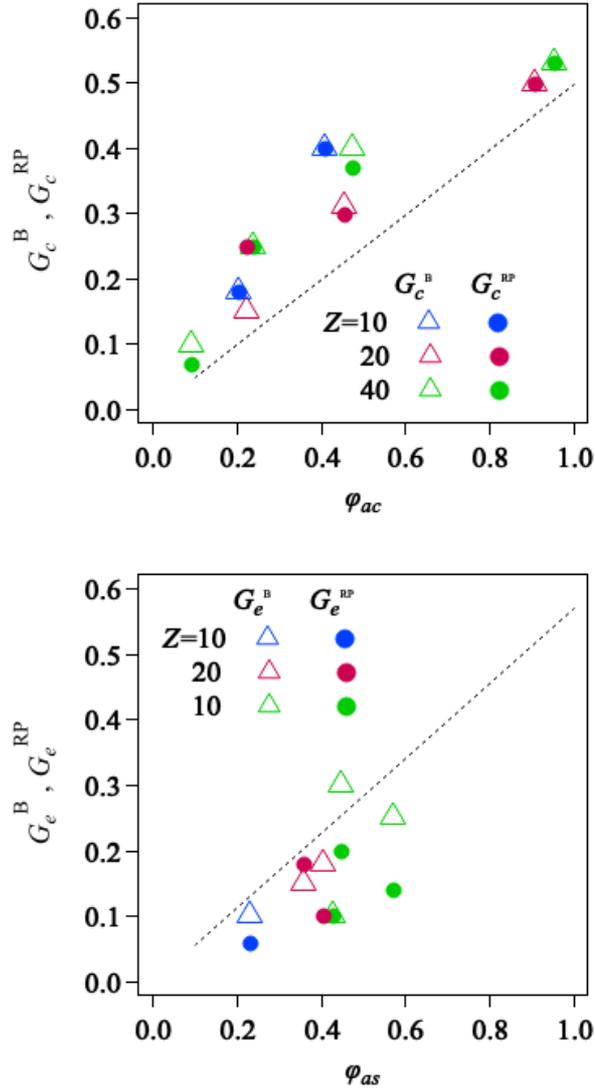

**Figure 6** Values of $G_c$ (top) and $G_e$ (bottom) used for the fitting in Figs 3 and 4 plotted against $\varphi_{ac}$ and $\varphi_{as}$ for $Z$=10 (blue), 20 (red) and 40 (green). Triangle and circle indicate the theories by Ball et al. [11] and by Rubinstein and Panyukov [12].

CONCLUSIONS

The primitive chain network simulations were conducted for polymer networks consisting of cross-links and slip-links. The examined systems were prepared from equilibrated entangled polymers by the random change of slip-links into cross-links. For the percolated networks, the stress-strain relationship was obtained and compared to the theories by Ball et al. [11] and Rubinstein and Panyukov[12]. Both theories were in reasonable agreement with the simulation results given that the moduli contributed from the cross-links, and slip-links were optimized. However, the parameter values are model-dependent and not consistent with the fraction of active links that carry the stress in the network.

The networks examined in this study are not ideal, in the sense that the clustering links hinder the chain sliding, and they disallow the system to reach the state of free-energy minimum. Besides, the non-Gaussian tension due to the numerical cut-off artificially raises the modulus. Nevertheless, the theories reasonably describe the stress-strain behavior of such systems by parameter tuning. This result reveals that the fitting of stress-strain data to the theories provides neither fraction of entanglement nor the amount of chain sliding at the entanglement, unless the examined network consists of Gaussian strands only and it correctly reaches the state of free-energy minimum.

ACKNOWLEDGEMENT

The authors thank the financial support from Ogasawara Foundation and JST-CREST (JPMJCR1992).

REFERENCES

1) Everaers R, *New J. Phys.*, **1**, 12.1-12.54 (1999).

2) Kuhn W, *J. Polym. Sci.*, **1**, 380–388 (1946).

3) Flory PJ, *Proc. R. Soc. London. A. Math. Phys. Sci.*, **351**, 351–380 (1976).

4) James HM, *J. Chem. Phys.*, **15**, 651 (1947).

5) James H, Guth E, *J. Polym. Sci.*, **4**, 153–182 (1949).

6) Ronca G, Allegra G, *J. Chem. Phys.*, **63**, 4990–4997 (1975).

7) Flory PJ, *J. Chem. Phys.*, **66**, 5720–5729 (1977).

8) Erman B, Flory PJ, *J. Chem. Phys.*, **68**, 5363–5369 (1977).

9) Edwards SF, *Proc. Phys. Soc.*, **92**, 9–16 (1967).

10) Rubinstein M, Panyukov S, *Macromolecules*, **30**, 8036–8044 (1997).

11) Ball RC, Doi M, Edwards SF, Warner M, *Polymer (Guildf).*, **22**, 1010–1018 (1981).

12) Rubinstein M, Panyukov S, *Macromolecules*, **35**, 6670–6686 (2002).

13) Edwards SF, Vilgis T a, *Reports Prog. Phys.*, **51**, 243–297 (1988).

14) Davidson JD, Goulbourne NC, *J. Mech. Phys. Solids*, **61**, 1784–1797 (2013).

15) Oberdisse J, Ianniruberto G, Greco F, Marrucci G, *EPL (Europhysics Lett.*, **58**, 530 (2002).

16) Oberdisse J, Ianniruberto G, Greco F, Marrucci G, *Rheol. Acta*, **46**, 95–109 (2006).

17) Takeda K, Sukumaran SK, Sugimoto M, Koyama K, Masubuchi Y, *Adv. Model. Simul. Eng. Sci.*, **2**, 11 (2015).

18) Masubuchi Y, Matsumiya Y, Watanabe H, Marrucci G, Ianniruberto G, *Macromolecules*, **47**, 3511–3519 (2014).

19) Yaoita T, Isaki T, Masubuchi Y, Watanabe H, Ianniruberto G, Marrucci G, *Macromolecules*, **44**, 9675–9682 (2011).

20) Yaoita T, Isaki T, Masubuchi Y, Watanabe H, Ianniruberto G, Marrucci G, *Macromolecules*, **45**, 2773–2782 (2012).

21) Masubuchi Y, Takimoto J-I, Koyama K, Ianniruberto G, Marrucci G, Greco F, *J. Chem. Phys.*, **115**, 4387–4394 (2001).

22) Masubuchi Y, *Annu. Rev. Chem. Biomol. Eng.*, **5**, 11–33 (2014).

23) Masubuchi Y, "Reference Module in Materials Science and Materials Engineering," Elsevier, 2016, pp.1–7.

24) Masubuchi Y, Ianniruberto G, Greco F, Marrucci G, *Rheol. Acta*, **46**, 297–303 (2006).

25) Masubuchi Y, Yaoita T, Matsumiya Y, Watanabe H, *J. Chem. Phys.*, **134**, 194905 (2011).

26) Masubuchi Y, Matsumiya Y, Watanabe H, Shiromoto S, Tsutsubuchi M, Togawa Y, *Rheol. Acta*, **51**, 1–8 (2012).

27) Masubuchi Y, Ianniruberto G, Marrucci G, *Soft Matter*, **16**, 1056–1065 (2020).

28) Masubuchi Y, Pandey A, Amamoto Y, Uneyama T, *J. Chem. Phys.*, **147**, 184903 (2017).

29) Furuichi K, Nonomura C, Masubuchi Y, Watanabe H, *J. Chem. Phys.*, **133**, 174902 (2010).

30) Mooney M, *J. Appl. Phys.*, **19**, 434–444 (1948).

31) Rivlin RS, *Philos. Trans. R. Soc. London. Ser. A, Math. Phys. Sci.*, **241**, 379–397 (1948).